\documentclass[%
 reprint,
 amsmath,amssymb,
 aps,
prb,
]{revtex4-1}

\usepackage{graphicx}
\usepackage{epstopdf}
\usepackage{dcolumn}
\usepackage{bm}
\usepackage{gensymb}

\begin{document}
\pdfoutput=1

\preprint{APS/123-QED}

\title{Quantum Oscillations in the Anomalous Spin Density Wave State of FeAs}%

\author{Daniel J. Campbell$^1$}
\author{Chris Eckberg$^1$}
\author{Kefeng Wang$^1$}
\author{Limin Wang$^1$}
\author{Halyna Hodovanets$^1$}
\author{Dave Graf$^2$}
\author{David Parker$^3$}
\author{Johnpierre Paglione$^1$}
\affiliation{$^1$Center for Nanophysics and Advanced Materials, Department of Physics, University of Maryland, College Park, Maryland 20742, USA}
\affiliation{$^2$National High Magnetic Field Laboratory, 1800 East Paul Dirac Drive, Tallahassee, Florida 32310, USA}
\affiliation{$^3$Oak Ridge National Laboratory, 1 Bethel Valley Road, Oak Ridge, Tennessee 37831, USA\\}
\date{\today}

\begin{abstract}

Quantum oscillations in the binary antiferromagnetic metal FeAs are presented and compared to theoretical predictions for the electronic band structure in the anomalous spin density wave state of this material. Demonstrating a new method for growing single crystals out of Bi flux, we utilize the highest quality FeAs to perform torque magnetometry experiments up to 35 T, using rotations of field angle in two planes to provide evidence for one electron and one hole band in the magnetically ordered state. The resulting picture agrees with previous experimental evidence for multiple carriers at low temperatures, but the exact Fermi surface shape differs from predictions, suggesting that correlations play a role in deviation from ab initio theory and cause up to a four-fold enhancement in the effective carrier mass. 

\end{abstract}

\maketitle

\section{\label{sec:Intro}Introduction}

The iron-based high-temperature superconductors, with transition temperatures reaching upwards of 60~K, are all comprised of a crystalline structure with layers of FeAs$_4$ tetrahedra\cite{FeSCsReview}. As a binary compound, FeAs itself naturally forms in an orthorhombic \textit{Pnma} MnP-type structure\cite{SelteFeAs} with a similar arrangement to the FeAs-based superconductors, except with octahedrally coordinated Fe atoms. Similar to both the parent compounds of the iron superconductors as well as the isostructural binaries CrAs\cite{WuCrAs, KotegawaCrAs} and MnP\cite{ChengMnP}, FeAs has an antiferromagnetic (AFM) ground state, with a N\'eel temperature T$_N$ = 70 K.\cite{SelteFeAs, FeAsPressure, SegawaFeAs, Gonzalez-AlvarezFeAs, SDWNeutron} However, unlike these other systems, to date FeAs has not been driven to a superconducting state by chemical substitution\cite{SelteVDoping, SelteCrDoping, IdoCrDoping} or pressure\cite{FeAsPressure}. This raises the question of how the electronic structure and/or magnetic interactions of FeAs set it apart from these other materials. 

Unlike CrAs and MnP, FeAs orders in a unique noncollinear spin density wave (SDW) state consisting of unequal moments along the \textit{a} and \textit{c}-axes with propagation along \textit{b}.\cite{SDWNeutron, FrawleyFeAs} Both spin amplitude and direction are modulated, and there is possible canting into the propagation direction. Despite a relatively extensive body of work on the properties of FeAs there is still uncertainty about the specifics of its electronic structure and what drives its magnetic order. Theoretical work by Parker and Mazin\cite{ParkerFeAsTheory} predicted the paramagnetic and AFM Fermi surfaces to differ substantially, with the AFM Fermi surface consisting of a single electron pocket at the $\Gamma$ point surrounded by four identical hole pockets. However, these and other calculations favor a more conventional AFM arrangement rather than the experimentally observed SDW.\cite{FrawleyFeAs, ParkerFeAsTheory, GriffinFeAs} Hall effect measurements have shown the coexistence of both hole and electron carriers over a wide temperature range\cite{SegawaFeAs,KhimFeAs,SaparovTAs}, but disagree over the dominant low temperature carrier.

In this paper we present a method to grow binary FeAs crystals using Bi flux, which produces samples with a larger residual resistivity ratio (RRR), defined as $\rho_{300~K}/\rho_{1.8~K}$, than the previously reported I$_2$ chemical vapor transport (CVT)\cite{SelteFeAs, SDWNeutron, FrawleyFeAs, SegawaFeAs, FeAsPressure, KhimFeAs} and Sn flux\cite{BlachowskiFeAs} techniques. A higher RRR is generally indicative of better crystal quality. Bi flux-grown crystals show quantum oscillations in magnetic torque measurements at high fields. Analysis of these oscillations makes it possible to give a more complete picture of the electronic structure of FeAs below T$_N$, allowing for comparison to previous theoretical and experimental results and the establishment of an experimentally verified Fermi surface in the SDW state.

\section{\label{sec:Exp. Details}Experimental Details}

In previous studies FeAs single crystals have been produced with I$_2$ as a transport agent\cite{SelteFeAs, SDWNeutron, FrawleyFeAs, SegawaFeAs, FeAsPressure, KhimFeAs} or Sn flux\cite{BlachowskiFeAs}. Being unsatisfied with the quality of crystals grown with these techniques, we explored alternative preparation methods. Ga, Zn, and Sb were unsuccessful as fluxes, but ultimately Bi was found to work well. Bi is advantageous because it does not form compounds with either Fe or As, reducing the possibility of impurity phases. At low temperatures, the resistivity $\rho$ is much smaller in Bi flux crystals, and accordingly we see a much higher RRR. RRRs consistently exceeded 70 with a maximum of 120, compared to 20\textendash{}40 with other growth methods\cite{SegawaFeAs, KhimFeAs}. Given that RRR is used as an indicator of crystal quality, we claim that Bi flux growth results in the highest quality FeAs single crystals yet produced.

To prepare the crystals, FeAs powder (either sintered in house or ground 99.5\% Testbourne pieces) was combined with polycrystalline Bi (Alfa Aesar Puratronic, 99.999\%) in a 1:20 ratio in an alumina crucible and sealed in a quartz ampule under partial Ar atmosphere. The growth was heated at 50~\degree{}C/hour to 900~\degree{}C, where it remained for two hours. The furnace was then cooled at a rate of 5~\degree{}C/hr to 500~\degree{}C, at which point the ampule was spun in a centrifuge to separate crystals from flux.

The crystal morphology when grown in this way is distinct from the polyhedral or platelike samples seen with CVT or Sn flux that we have also made. Crystals grown in Bi flux are needlelike (Fig. 1b, inset), with typical dimensions of 0.03~$\times$~0.03~$\times$~0.8~mm$^3$. Powder X-ray diffraction measurements using a Rigaku MiniFlex600 give lattice parameters in the \textit{Pnma} space group as \textit{a}~=~5.44 \AA{}, \textit{b}~=~6.02~\AA{}, and \textit{c}~=~3.37 \AA{}, in line with previous results (note that axis conventions differ between papers)\cite{SelteFeAs, SegawaFeAs, SDWNeutron}. The long direction of the crystal was always the \textit{c}-axis, as verified by Laue photography and single crystal XRD and inferred from the initial increase in resistivity with decreasing temperature that is unique to measurement along [001]\cite{SegawaFeAs, KhimFeAs}. For the sample used in oscillations measurements, the orientation of the \textit{a}-axis was similarly confirmed with XRD and Laue, making the \textit{b}-axis the remaining perpendicular direction. Composition was confirmed by energy dispersive X-ray spectroscopy as almost exactly 1:1 for a large number of samples from different growths. There was no sign of Bi contamination in EDS, XRD, or transport measurements. One drawback of the Bi flux growth method is that the small, thin samples are ill-suited for Hall effect or single crystal susceptibility measurements, as they are too light and narrow.

Electrical resistivity measurements were performed down to 1.8~K in a 9~T and 14~T Quantum Design Physical Properties Measurement System (PPMS). Measurements of magnetization using torque cantilevers and electrical transport were made at the DC Field Facility of the National High Magnetic Field Laboratory (NHMFL) in Tallahassee, Florida. A He-3 system with a base temperature of about 350~mK was used in both the 31~T, 50~mm bore and 35~T, 32~mm bore magnets. Measurements were conducted up to 31.5~T for both resistance and torque and 35~T for torque alone.

\section{\label{sec:Transport}Transport Results}

\begin{figure}
    \centering
    \includegraphics[width=0.42\textwidth]{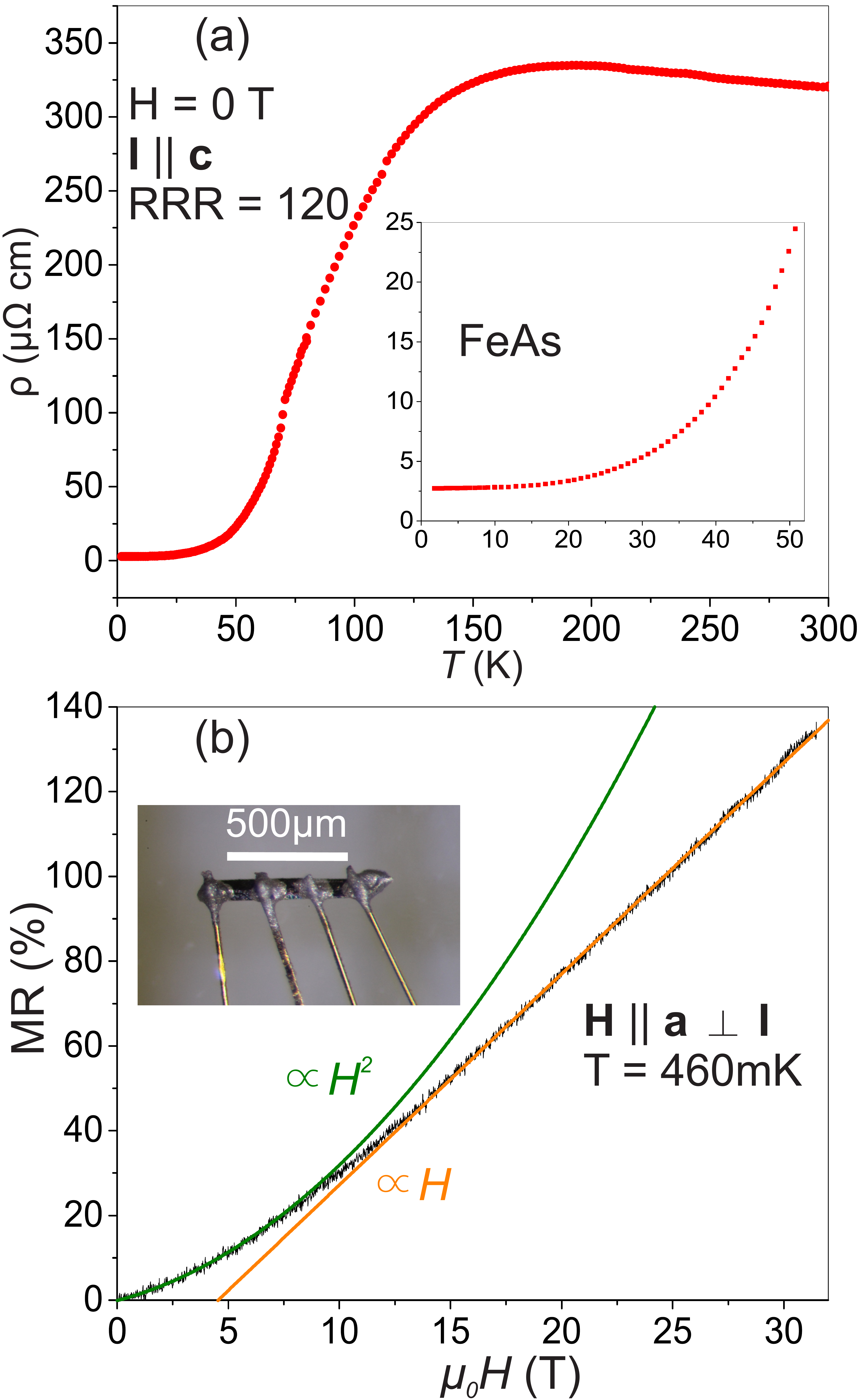}
    \caption{(a) Resistivity vs. temperature for an FeAs crystal grown from Bi flux. The high RRR and low residual resistivity (inset) indicate very good crystal quality. (b) Magnetoresistance data as a percentage of 0~T resistivity for FeAs up to 31.5~T. Fits of low and high field data to quadratic and linear functions, respectively, show a transition in field dependence of MR around 10~T. Inset: an FeAs crystal wired for longitudinal resistance measurements, showing the needlelike geometry particular to Bi flux growth.}
    \label{fig:Figure1}
\end{figure}

The temperature-dependent resistivity for single crystal FeAs is shown in Fig. 1. The 300~K resistivity value for Bi flux crystals is about 300~$\mu\Omega$~cm, similar to what has been seen previously\cite{SegawaFeAs, KhimFeAs}. As those studies note, an initial increase in $\rho{}$ as temperature decreases, with a maximum near 150~K, signifies that the measurement is conducted with \textbf{I} $\parallel$ \textbf{c}. A kink at 70~K marks the SDW onset at the same temperature as in other transport, susceptibility, and heat capacity measurements\cite{SelteFeAs, FeAsPressure, SegawaFeAs, Gonzalez-AlvarezFeAs}. The inset to Fig. 1a shows the resistivity plateauing below 20 K at about 2.5~$\mu\Omega$~cm.

With increasing field, magnetoresistance (MR) in FeAs evolves from the typical metallic \textit{H$^2$} dependence to being linear in \textit{H} (Fig. 1b). Linearity continues without saturation up to 31.5~T. This quadratic-to-linear crossover has previously been reported to occur at about 6 T for measurements at 10 K.\cite{KhimFeAs} Our samples show it occurring at roughly 10 T below 1 K with \textbf{H} $\perp$ \textbf{I}. Arsenic vacancies have been identified as sources of disorder leading to linear MR in other compounds.\cite{NarayananCd3As2} Additionally, some lower RRR Bi flux-grown samples showed a low temperature upturn in resistivity, which in other layered systems whose structures contain As ``nets'' has been linked to As vacancies.\cite{CichorekAsVacancies} It is possible that the presence of As vacancies in FeAs crystals depends on the growth method and affects transport properties, although no sign of As deficiency was seen in EDS. The data in Fig. 1b may show the onset of oscillatory behavior just below 30~T. However, there was not enough data for possible analysis and no sign of oscillations appeared upon rotating the sample.

\section{\label{sec:QOs}Quantum Oscillations}

Quantum oscillations arise when a material reveals its band structure in the presence of a magnetic field, forming quantized Landau levels whose spacing is proportional to field strength. Changing field causes these bands to pass through the chemical potential, and the resulting change in occupancy produces an oscillatory signal that can be detected in a wide variety of density of states-dependent quantities, most commonly resistance (in which case they are called Shubnikov-de Haas oscillations) and magnetization (called de Haas-van Alphen oscillations).\cite{ShoenbergOscillations, CarringtonQOs, VignolleQOs} For single crystal FeAs, torque data show multiple frequencies across different angles of applied field, as evident in Fig. 2a which shows the raw torque signal at several field orientations. Oscillatory behavior was clear in the torque signal as low as 10 T at some angles.

\begin{figure}[t]
    \centering
    \includegraphics[width=0.45\textwidth]{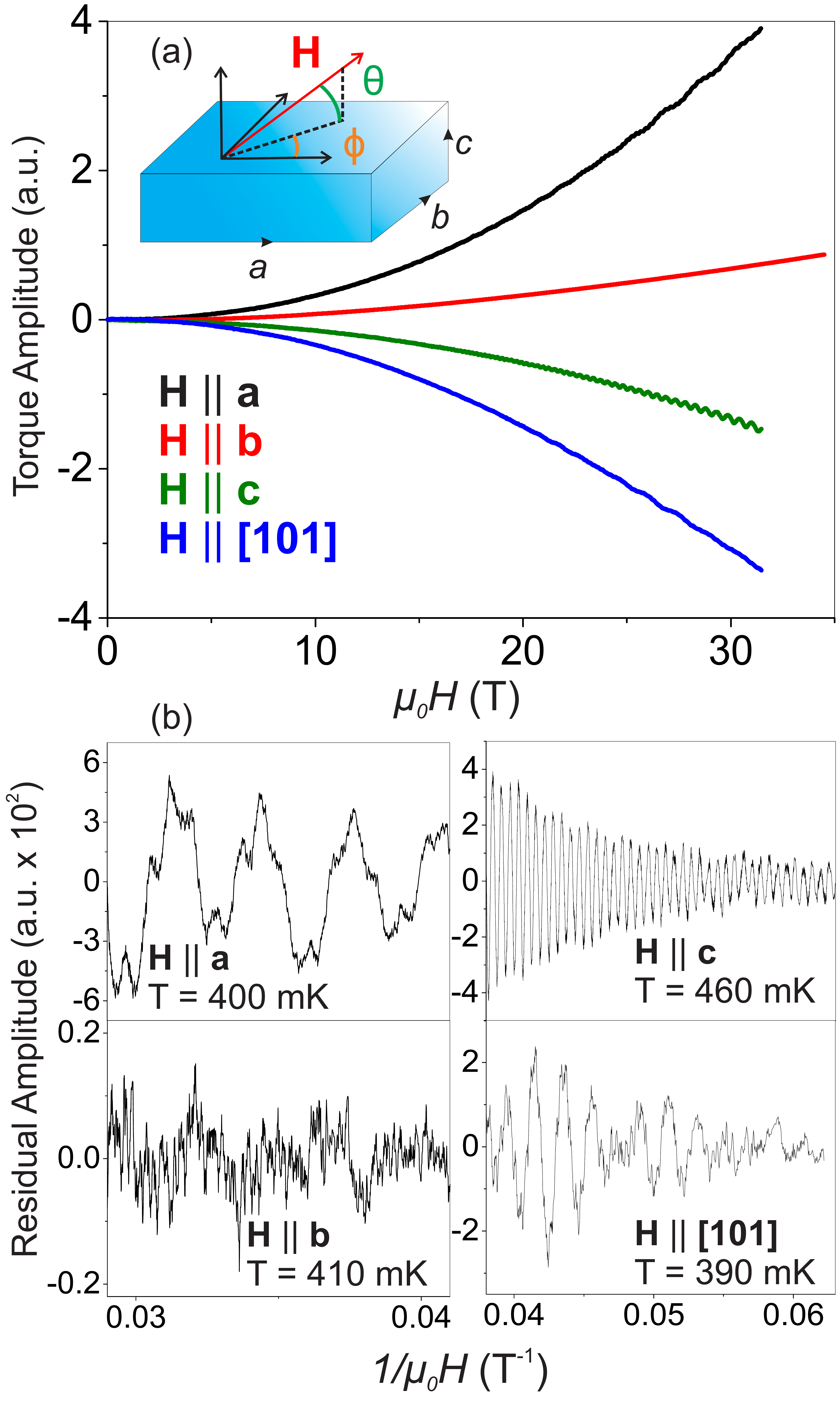}
    \caption{(a) Raw torque cantilever data for FeAs for several field orientations. Inset: a schematic showing how field angle was changed in the two measurements. Measurements to 31.5~T went from \textbf{H} $\parallel$ \textbf{a} to \textbf{c} ($\phi{}$~=~0\degree{}, sweeping $\theta{}$). Those to 35~T went from \textbf{H}~$\parallel$~\textbf{a} to \textbf{b} ($\theta{}$~=~0\degree{}, sweeping $\phi{}$). (b) The residual oscillatory signal of the raw data. Amplitudes are arbitrary but consistent relative to those in (a), and have been enhanced by a factor of 100.}
    \label{fig:Figure2}
\end{figure}

Two sets of measurements were made on the same crystal as it was rotated in two different planes relative to magnetic field, as illustrated in the schematic in Fig. 2a. Data in the 31~T magnet were taken at 24 angles with $\phi$~=~0\degree{} and changing $\theta{}$. In this configuration $\theta{}$~=~0\degree{} signifies \textbf{H}~$\parallel$~\textbf{a} and $\theta$~=~90\degree{} is \textbf{H}~$\parallel$~\textbf{c}. Up to 35~T, 16 measurements were made with $\theta{}$ kept at 0\degree{} while $\phi$ was changed, corresponding to \textbf{H}~$\parallel$~\textbf{a} at $\phi$~=~0\degree{} and \textbf{H}~$\parallel$~\textbf{b} at $\phi$~=~90\degree{}.

To extract the oscillatory component a 3rd order polynomial was subtracted from the raw data; Fig. 2b shows examples of the presence of different frequencies at different angles, and changes in the amplitude of the residual torque signal. Fast Fourier transforms (FFT) were then performed on the residual data to obtain a frequency spectrum (Fig. 3a and 3b).

\subsection{\label{sec:Angle}Angular Dependence}

\begin{figure}
    \centering
    \includegraphics[width=0.45\textwidth]{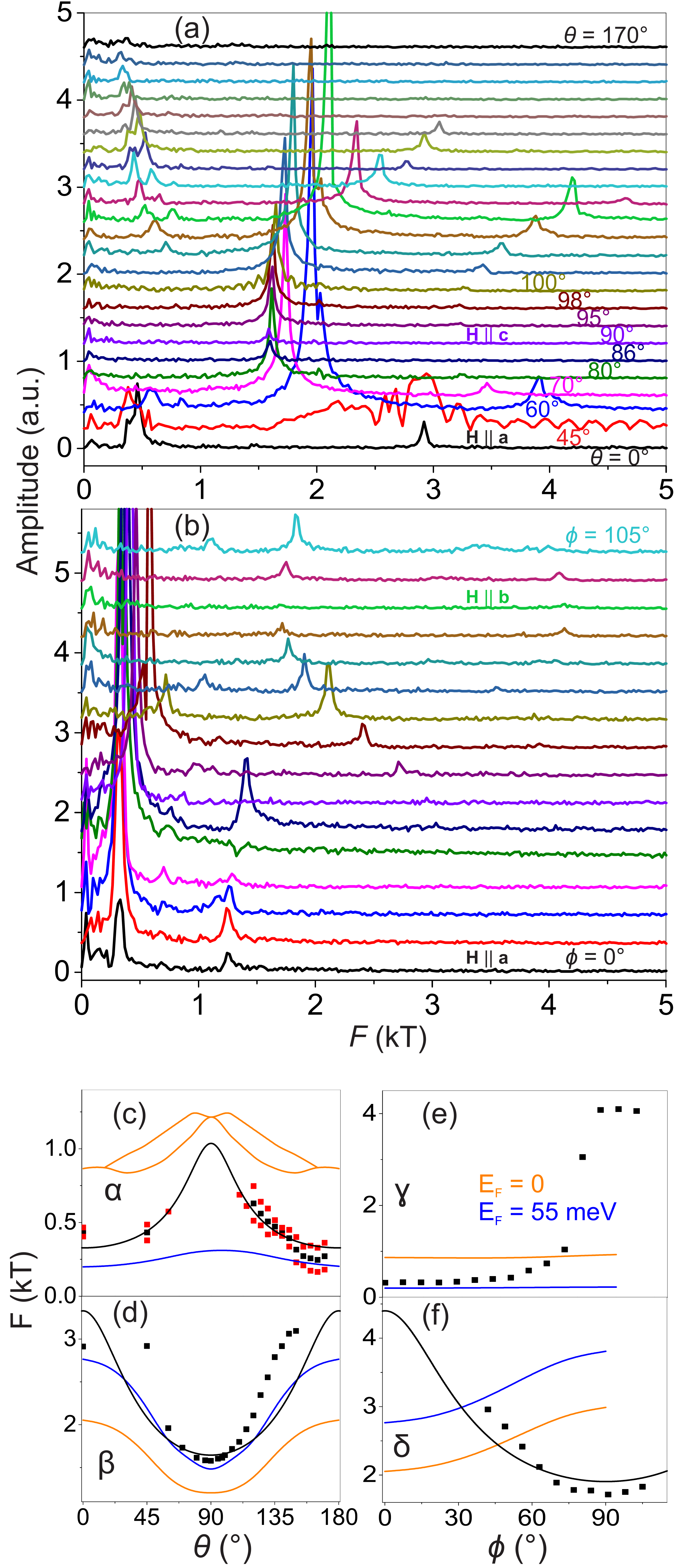}
    \caption{FFTs at base temperature (350\textendash{}550~mK) of oscillatory signals for all angles, offset for clarity. In (a) field goes from parallel to \textbf{a} to parallel to \textbf{c}, changing $\theta{}$ (in 5\degree{} increments at higher angles). In (b) $\phi{}$ is swept (in 7\degree{} increments over the entire range). (c)-(f) show observed peaks for each oscillation band as a function of angle as well as theoretically generated frequencies based on AFM Fermi surface calculation of Parker and Mazin\cite{ParkerFeAsTheory} (orange line) and the same calculation with E$_F$ raised by 55~meV (blue). For $\alpha{}$, $\beta{}$, and $\gamma{}$ fits to a perfectly ellipsoidal Fermi surface are also given (black lines). For (c), where peak splitting occurred the average (black) of the $\alpha_1$ and $\alpha_2$ frequency peaks (red) was used for the fit.}
    \label{fig:Figure3}
\end{figure}

By plotting FFT data for all angles of the two runs as in Figs. 3a and 3b, it is clear that although frequency values vary substantially when sweeping either $\theta{}$ or $\phi{}$, they correspond to one of five extremal Fermi Surface orbits. Harmonics of these five frequencies also appear at integer multiples. In the $\theta$-scan (Fig. 3a, \textbf{H} in the \textit{ac} plane) two low frequencies (denoted $\alpha{}_1$ and $\alpha{}_2$) around 500~T, and one higher frequency peak near 1.5~kT ($\beta{}$), were observed. The proximity of the two $\alpha{}_i$ peaks indicates that they arise from the same Fermi surface pocket, with two slightly displaced extremal orbits. The $\alpha_1$\textendash{}$\alpha_2$ frequency difference was roughly 150~T, independent of temperature or angle.

For the measurement varying $\phi$ (Fig. 3b, \textbf{H} in the \textit{ab} plane) we see two peaks: one with a frequency of about 300~T for angles closer to 0\degree{} ($\gamma{}$) and a higher frequency peak with F $\approx$ 2~kT ($\delta{}$). However, the $\gamma{}$ peak diverged to much higher frequencies exceeding $\delta$ near \textbf{H}~$\parallel$~\textbf{b} with a substantially reduced amplitude. As Fig.~3b shows, the amplitude decreased substantially as this change occured. It is notable that this divergence comes for \textbf{H}~$\parallel$~\textbf{b}, as that is the SDW propagation direction\cite{SDWNeutron, FrawleyFeAs} and the field direction for which no kink is observed in susceptibility at T$_N$\cite{SegawaFeAs}.

From seeing two main orbits (one of which shows some frequency splitting) in both the \textit{a}\textendash{}\textit{b} and the \textit{a}\textendash{}\textit{c} field rotation studies, we can conclude that there are two distinct pockets of the Fermi surface giving rise to extremal orbits that produce the observed dHvA oscillations. This fits with the theoretical prediction of one unique electron and hole pocket each in the magnetic state\cite{ParkerFeAsTheory} as well as experimental evidence suggesting multiple carriers in this regime\cite{SegawaFeAs, KhimFeAs, SaparovTAs}.

\subsection{\label{sec:Fermi}Fermi Surface Shape}

The angular dependence of peak frequency can be used to model the shape of the Fermi surface.\cite{ShoenbergOscillations} Specifically, for an ellipsoidal Fermi surface the frequency should vary with angle as $F~=~\frac{F_0}{\sqrt{cos^2\psi~+~\frac{1}{\epsilon}sin^2\psi}}$, where $F_0$ is the maximum frequency, $\psi$ the angle and $\epsilon$ the eccentricity of the cross sectional ellipse in the plane of rotation.\cite{SebastianOscns} Figs.~3c, 3d, and 3f show fits of peak frequency to this equation for $\alpha{}$, $\beta{}$, and $\delta{}$. At angles with split $\alpha{}_1$ and $\alpha{}_2$ frequencies, their average value was used for the fit. Divergence from fits makes it clear that the pockets are not perfectly ellipsoidal, however the qualitative agreement shows that $\alpha$, $\beta$, and $\delta$ correspond to orbits around three dimensional parts of the Fermi surface with a generally ellipsoidal shape. In contrast, $\gamma{}$ shows a slight increase in frequency at lower angles, until roughly $\phi$~=~70\degree{} when frequency increases by an order of magnitude before plateauing. This behavior is closer to that of cylindrical or two dimensional pockets, although $\gamma{}$ does not fit well to the inverse cosine dependence expected from a perfect cylinder.

Fig. 4 shows two theoretical Fermi surfaces for antiferromagnetic FeAs obtained with density functional theory (DFT). Calculations were done for the ``AF2'' state, calculated by Parker and Mazin to be most favorable at low temperatures\cite{ParkerFeAsTheory}, in which Fe atoms align antiferromagnetically with both nearest and next-nearest neighbors. This same arrangement was  favored in the calculations of Frawley et al.\cite{FrawleyFeAs}, whereas Griffin and Spaldin\cite{GriffinFeAs} differed in having a ferromagnetic arrangement of next-nearest neighbors. Neither of these orderings matches the SDW. The top surface in Fig. 4 uses the original AF2 Fermi level, while the bottom one is from the same calculation but with the Fermi level raised by 55~meV. This shift changes the size of the pockets, but their shapes and locations are unaltered, establishing the robustness of this Fermi surface geometry and therefore also of the expected angular dependence of oscillation frequencies. In either case there is an electron pocket at the central $\Gamma{}$ point and four identical hole pockets at ($k_a$, $k_b$, $k_c$) = ($\pm{}$0.25, $\pm{}$0.3, 0).

Theoretical quantum oscillation frequencies generated from the DFT calculations using the Supercell K-space Extremal Area Finder (SKEAF) program\cite{RourkeSKEAF} are plotted together with the experimental data in Figs. 3c-3f. Two bands, one electron-like and one hole-like, were expected for each plane of field rotation, a 1:1 correspondence to what was obtained in measurements. Based on expected frequencies and angular dependence we were able to identify the $\alpha$ and $\gamma$ peaks as hole pocket oscillations, with $\beta$ and $\delta$ belonging to the electron band. Increasing the Fermi level does not change the angular dependence, but the change in pocket size gives closer agreement to the observed oscillation frequencies (which are proportional to the cross sectional pocket area) in most cases. For $\alpha{}$ and $\beta{}$ expected angular dependence matches well to data, and in fact the splitting seen in the hole band is also present in the unshifted Fermi surface calculation in the range $\theta$~=~30\degree{}\textendash{}90\degree{}. This reinforces the roughly ellipsoidal pockets inferred from experimental angular dependence. For $\gamma{}$ the divergence at higher angles does not happen in the theory, where frequency has a much smaller expected increase. For $\delta{}$ a variation of frequency with angle is seen, however the locations of the maximum and minimum oscillation frequencies are reversed. This indicates that the electron pocket area is larger in the $k_b$\textendash{}$k_c$ plane than in the $k_a$\textendash{}$k_c$ plane, the opposite of the band structure prediction. Overall the DFT Fermi surface appears to give an accurate description of electron and hole pocket shape for field rotated between the \textit{a} and \textit{c}-axes ($\phi$~=~0\degree{}, changing $\theta$), but not the \textit{a} and \textit{b}-axes ($\theta$~=~0\degree{}, changing $\phi$). Again we note that $k_b$ corresponds to the propagation direction of the SDW, while the moments lie in the $k_a$\textendash{}$k_c$ plane. The fact that this is also the field direction for which we see the strongest divergence from calculation points to a connection between disagreement of DFT and experiment over both magnetic ordering (as was already known) and band structure (as we have shown here).

Oscillations data do, however, support the two carrier picture put forth by other groups\cite{SegawaFeAs, KhimFeAs, SaparovTAs}. We have established that the SDW Fermi surface geometry is roughly ellipsoidal, in which case carrier concentration can be calculated from oscillation frequency F as $n~=~\frac{1}{3\pi{}^{2}}(\frac{2eF}{\hbar})^{\frac{3}{2}}$ where $e$ is the electron charge and $\hbar$ the reduced Planck constant.\cite{NarayananCd3As2} Applying this equation to our data gives ranges of 2.2~$\times$~10$^{19}$~\textendash{}~1.5~$\times$~10$^{21}$~cm$^{-3}$ for the hole pocket and 3.5~\textendash{}~9.6~$\times$~10$^{20}$~cm$^{-3}$ for the electron pocket, based on maximum and minimum observed frequencies. These are slightly different than the values n$_h$~=~8~$\times$~10$^{18}$~cm$^{-3}$ and n$_e$~=~1~$\times$~10$^{21}$~cm$^{-3}$ found by Khim et al.\cite{KhimFeAs} through a fit of MR data. The hole pocket has a much more dramatic angular dependence, and for a small range of angles near \textbf{H}~$\parallel$~\textbf{b} even exceeds the electron value. Assuming comparable scattering rates, this anisotropy could account for the sign change in R$_H$ at low temperature seen by Segawa and Ando\cite{SegawaFeAs} but not Khim et al.\cite{KhimFeAs} Hence it is possible that the dominant carrier in transport measurements of FeAs depends on the direction of applied current.

The area of a cyclotron orbit can be calculated directly from the oscillation frequency using the Onsager formula $A = \frac{2\pi{}eF}{\hbar}$.\cite{ShoenbergOscillations} For \textbf{H}~$\parallel$~\textbf{a} the hole oscillation frequency is 316~T and its cyclotron orbit covers about 4~nm$^{-2}$ or 2\% of the \textit{k$_b$\textendash{}k$_c$} first Brillouin zone. The oscillation is not observed for \textbf{H}~$\parallel$~\textbf{c}, but for \textbf{H}~$\parallel$~\textbf{b} it increases substantially to 39~nm$^{-2}$ or 33\% of the first Brillouin zone. The electron pocket shows less angular dependence. For \textbf{H}~$\parallel$~\textbf{a}, \textbf{b}, and \textbf{c} the electron orbit covers an area of 28, 16, and 15~nm$^{-2}$, respectively, which in each case corresponds to very nearly 13\% of the in-plane area of the first Brillouin zone. Thus while the electron pocket still has a three dimensional shape it takes up the same proportion of the Fermi surface along principal axes, in contrast to highly anisotropic behavior in the hole band. 

\subsection{\label{sec:LK}Temperature Dependence}

Tracking the decrease in oscillation amplitude with increasing temperature gives an estimate of effective mass through the Lifshitz-Kosevich (LK) factor $R_T~=~\frac{\alpha{}m\text{*}T/(\mu{}_0Hm_e)}{\text{sinh}(\alpha{}m\text{*}T/(\mu{}_0Hm_e))}$ where \textit{m*} is the effective carrier mass, $m_e$ the electron mass, $\mu{}_0$ the vacuum permeability, and $\alpha{}~=~2\pi^2ck_B/e\hbar{}~\approx 14.69$ T/K with \textit{c} the speed of light and $k_B$ the Boltzmann constant.\cite{ShoenbergOscillations} Temperature dependence was taken at three field orientations: $\theta$~=~$\phi$~= 0\degree{} (\textbf{H}~$\parallel$~\textbf{a}), $\phi$~=~0\degree{}, $\theta$~=~98\degree{} (near \textbf{H}~$\parallel$~\textbf{c}) and $\phi$~=~0\degree{}, $\theta$~=~135\degree{} (\textbf{H} halfway between the \textit{a} and \textit{c}-axes). Oscillatory signals for these orientations are shown in Fig. 2b. The second angle gives an idea of the effective mass along the \textit{c}-axis, but $\theta$ was not set to exactly 90\degree{} since the torque signal was much reduced directly along that axis. Fig. 5a gives an example of the clear suppression of FFT amplitude of $\alpha_1$, $\alpha_2$, and $\beta$ with temperature for \textbf{H}~$\parallel$~\textbf{[101]}.

\begin{figure}[t]
    \centering
    \includegraphics[width=.47\textwidth]{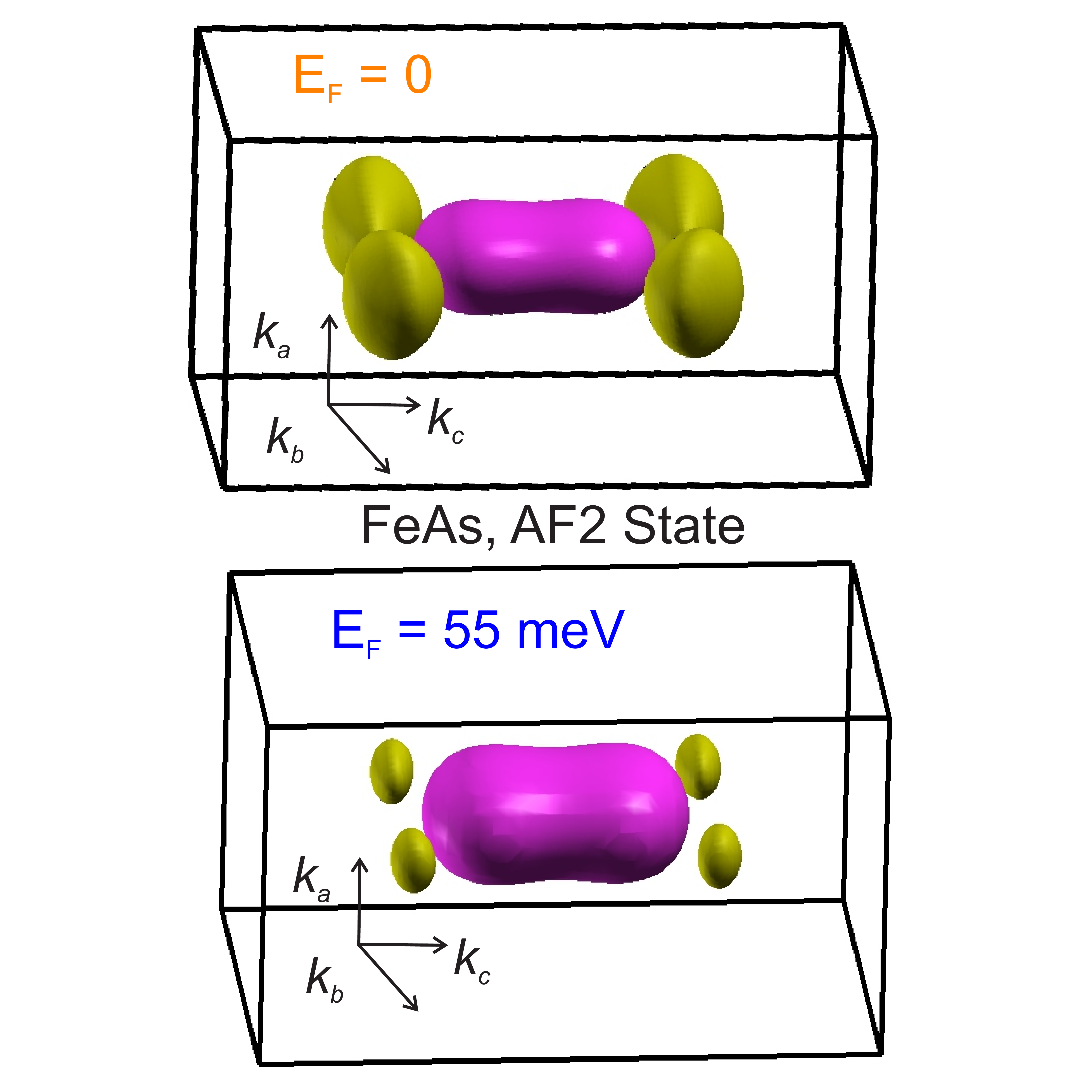}
    \caption{DFT calculated Fermi surfaces of FeAs in the predicted ``AF2'' magnetic state\citep{ParkerFeAsTheory}, consisting of an electron pocket (pink) at the $\Gamma$ point and four identical, symmetrically oriented hole pockets (yellow). The bottom has had the Fermi energy raised by 55~meV, which changes the size of the pockets but not their location or overall shape.}
    \label{fig:Figure4}
\end{figure}

Temperature dependent amplitudes for three different field orientations are shown in Fig. 5b-d. Table I gives the extracted effective masses. With \textbf{H}~$\parallel$~\textbf{a} (Fig. 5b), only the $\gamma{}$ hole pocket (F = 315 T) is seen. As with oscillation frequencies, we can compare experimental effective masses to those generated from the DFT Fermi surface for the original or 55~meV shifted Fermi level.\cite{RourkeSKEAF} A fit to the LK equation at this angle gives an effective mass of 3.1\textit{m$_e$}, larger than the theoretical predictions of 1.78\textit{m$_e$} (E$_F$~=~0~eV) and 1.138\textit{m$_e$} (E$_F$~=~55~meV) for the hole pocket in the same orientation. For $\theta$~=~98\degree{} (Fig. 5c) only $\beta{}$, at 1.61~kT, appears. Given the absence of any other frequencies, amplitude can be directly extracted from the oscillatory data, and the effective mass is \textit{m*}~=~1.2\textit{m$_e$}. For $\theta$~=~90\degree{} the predicted electron band masses are 0.668\textit{m$_e$} and 0.812\textit{m$_e$} (0~eV) or 0.6322 and 0.940\textit{m$_e$} (55~meV). At 135\degree{} (Fig. 5d), $\beta{}$ (now at 2.77~kT) survives only up to 1.8~K. Due to the presence of the lower $\alpha$ frequencies in the residual signal, amplitude is taken from the FFT. The LK fit gives \textit{m*}~=~3.2\textit{m$_e$}, nearly a factor of three larger than its value at $\theta$~=~98\degree{}. This again exceeds predictions of 1.124\textit{m$_e$} (0~eV) or 1.252\textit{m$_e$} (55~meV).

At $\theta$~=~135\degree{}, $\alpha{}_1$ and $\alpha{}_2$ are found at 412 and 536~T. Using the average value of the amplitude of the two peaks we find an effective mass of $m_{\alpha{},~ave}$~=~3.8$m_e$. The individual peaks have similar values, further supporting the idea that they arise from the same band. This number is similar to the value of 3.1$m_e$ obtained for the same pocket for \textbf{H}~$\parallel$~\textbf{a}. The 0 eV Fermi level prediction is for two peaks with masses 1.561\textit{m$_e$} and 2.023\textit{m$_e$}, while that for 55~meV is one peak of 1.301\textit{m$_e$}. As with all other measured angles, the experimental effective masses are larger than those predicted.

The general trend of enhanced experimental effective masses indicates the presence of correlation effects unaccounted for by DFT. It has recently been proposed that spin-orbit coupling may have significant influence on the FeAs band structure in the magnetic state,\cite{FrawleyFeAs} even though it is not normally included in calculations for Fe-based compounds. However, this and other correlated effects may account for some of the disagreement we observe between theory and experiment.

\begin{figure}
    \centering
    \includegraphics[width=0.48\textwidth]{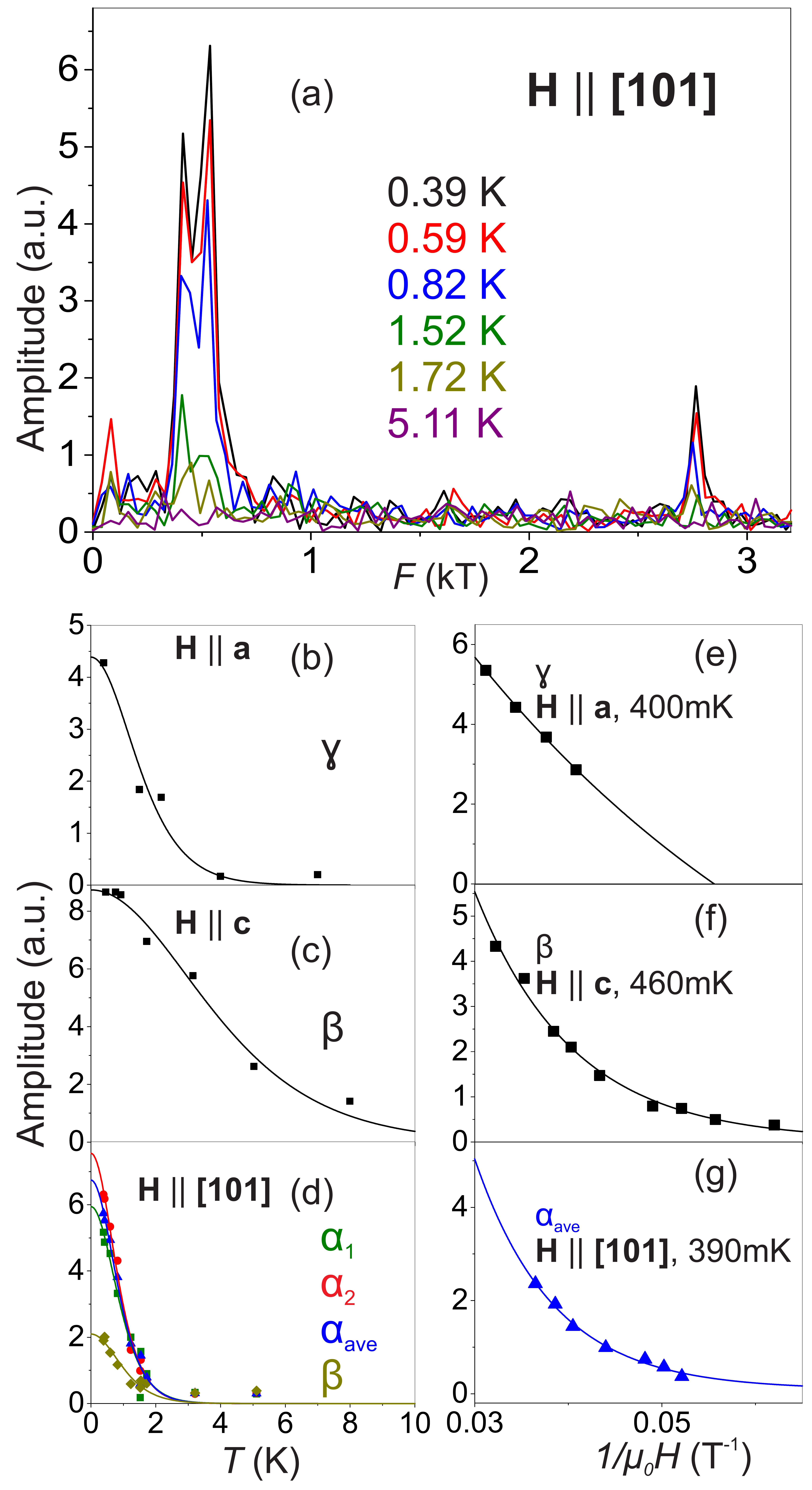}
    \caption{(a) FFT at multiple temperatures for \textbf{H}~$\parallel$~\textbf{[101]}, showing the decrease in amplitude with temperature of the $\alpha$ (0.5~kT) and $\beta$ (2.8~kT) peaks. (b-d) Fits to the LK formula at different field angles. Plots of peak amplitudes vs. inverse field at base temperature (e-g) were fit to the Dingle formula. Base temperature varied slightly between measurements. Due to its small amplitude relative to $\alpha_1$ and $\alpha_2$, it was not possible to extract T$_D$ for $\beta{}$ at \textbf{H}~$\parallel$~\textbf{[101]}.}
    \label{fig:Figure5}
\end{figure}

\subsection{\label{sec:Dingle}Dingle Temperature and Scattering}

The Dingle factor in the oscillation amplitude is $R_D~=~\text{exp}(-\alpha{}m\mbox{*}T_D/(\mu{}_0Hm_e))$, where $\alpha~\approx{}~14.69$~T/K again and T$_D$ is the Dingle temperature.\cite{ShoenbergOscillations} T$_D$ is proportional to the scattering rate $\Gamma$ as T$_D = \frac{\hbar}{2\pi{}k_B}\Gamma$. Dingle temperatures were calculated based on fits of peak amplitudes versus inverse field (Figs. 5e-5g) and are listed in Table I along with $\Gamma$ values. It is only possible to solve for T$_D$ if the effective mass is known, limiting the analysis to only the three angles for which temperature dependent measurements were made. Additionally, since calculation of T$_D$ relies on a clear exponential decay of amplitude it is typically necessary to have one dominant peak at a specific angle to extract a Dingle temperature. For that reason we could not calculate T$_D$ for each peak at each angle. This is not an issue for \textbf{H}~$\parallel$~\textbf{c}, where only $\beta$ appears and T$_D$~=~5.5~K. However, due to their very similar frequencies it is hard to separate $\alpha_1$ and $\alpha_2$, and so we can only give T$_D$~=~2.2~K for the average $\alpha$ oscillation at $\theta$ = 135\degree{}. The $\beta$ oscillation is only a small modulation of the signal for this orientation (Fig. 2c). Again there is anisotropy in the hole pocket, as T$_D$ goes from 2.2~K to 5.1~K as field moves from [101] to the \textit{a}-axis. This is not surprising given the previously noted differences in transport data for measurements along different crystal axes.\cite{SegawaFeAs}

Comparing binary FeAs to the iron pnictide superconductors, the Dingle temperatures we observe are similar to those seen in BaFe$_2$As$_2$. In that material T$_D$ can be calculated for two out of three observed bands, and for both is in the range 3\textemdash{}4~K.\cite{AnalytisBa122QOs} In another 122 material, KFe$_2$As$_2$, T$_D$ is between 0.1\textemdash{}0.2~K for five different pockets.\cite{TerashimaK122QOs} For the K compound, RRR values up to 2000 are possible\cite{NakajimaK122RRR}, while for BaFe$_2$As$_2$ RRR less than 10 is typical\cite{SefatBa122RRR}, though it can be raised to nearly 40 with annealing\cite{RotunduBa122Anneal}. Despite a higher RRR, scattering rates in FeAs are on the same level as those in BaFe$_2$As$_2$ rather than a ``cleaner'' material like KFe$_2$As$_2$.

\begin{table}
	\centering

    \caption{Parameters extracted from fits of FeAs quantum oscillation amplitude to the LK and Dingle factors at several field orientations.\\}
    \label{tab:Table1}

\renewcommand{\arraystretch}{1.2}
\begin{tabular}{ | c | c | c | c | c | c | c | c |}
	\hline
	Orbit & Type & \textbf{H} $\parallel$ \textbf{[hkl]} & F (T) & \textit{m*}/$m_e$ & T$_D$ (K) & $\Gamma$ (10$^{12}$ s$^{-1}$) \\
	\hline
	$\alpha_1$ & h & \textbf{[101]} & 412 & 3.6 & \textemdash{} & \textemdash{} \\
	\hline
	$\alpha_2$ & h & \textbf{[101]} & 536 & 3.9 & \textemdash{} & \textemdash{} \\
	\hline
	$\alpha_{ave}$ & h & \textbf{[101]} & \textemdash{} & 3.8 & 2.2 & 1.8 \\
	\hline
	$\gamma$ & h & \textbf{[100]} & 316 & 3.1 & 5.1 & 4.2 \\
	\hline
	$\beta$ & e & \textbf{[101]} & 2765 & 3.2 & \textemdash & \textemdash \\
	\hline
	$\beta$ & e & \textbf{[001]} & 1615 & 1.2 & 5.5 & 4.5 \\
	\hline
\end{tabular}

\end{table}

\section{\label{sec:Conclusion}Conclusion}

Growing FeAs (T$_N$~=~70 K) out of Bi flux has proven to produce higher quality crystals than previous attempts with Sn flux or I$_2$ CVT. This improved quality has made it possible to observe quantum oscillations in the torque signal at high field. Measurements in two different planes reveal five unique peaks, corresponding to one electron and one hole band in each direction (with the hole band split for field in the \textit{ac} plane). These peaks can be indexed using a DFT-calculated Fermi surface for antiferromagnetic FeAs.\cite{ParkerFeAsTheory} Three peaks near 500~T (split peaks $\alpha_1$ and $\alpha_2$ and $\gamma{}$) stem from extremal orbits around the predicted four identical hole pockets, and two others ($\beta$ and $\delta$, one in each plane) near 2~kT come from the electron pocket at the $\Gamma{}$ point. The $\gamma{}$ oscillation band has a two dimensional shape and cannot be easily assigned a simple geometry. The other three observed oscillations show a three dimensional, qualitatively ellipsoidal angular dependence, as expected from calculations, with slight disagreement in pocket size.

The observation of two distinct frequencies overall validates the multiband notion of transport in the low temperature SDW state indicated by previous experiment.\cite{SegawaFeAs, KhimFeAs, SaparovTAs} We see good agreement with the calculated Fermi surface when field is swept in the \textit{ac} plane, but disagreement for \textit{ab} plane rotation. Most notable is a large increase in the cross sectional area of the hole pocket near the $k_a$\textendash{}$k_c$ plane, where it becomes larger than the electron pocket. Extracted effective masses for both hole and electron pockets are enhanced over predictions, indicating the likely presence of correlated electron effects. It was already known that theoretical calculations did not match the magnetic state of FeAs, and through quantum oscillations measurements we have shown that the band structure also awaits a full theoretical description.

\section{\label{sec:Acknowledge}Acknowledgments}

We wish to thank Nick Butch for assistance with Laue photography at the NIST Center for Neutron Research. This research was supported by Air Force Office of Scientific Research award No. FA9550-14-1-0332 and National Science Foundation Division of Materials Research award No. DMR-1610349. A portion of this work was performed at the National High Magnetic Field Laboratory, which is supported by National Science Foundation Cooperative Agreement No. DMR-1157490 and the State of Florida. We acknowledge the support of the Maryland NanoCenter and its FabLab.

\bibliography{FeAsRefs}

\end{document}